\documentstyle[12pt]{article}
\begin{document}

\title{I am the cat who walks by himself}

\author{{Asher Peres}%
\footnote{Department of Physics, Technion---Israel Institute of
Technology, 32000 Haifa, Israel.}}\date{}

\maketitle
\begin{abstract} 
\noindent The city of lions. Beaulieu-sur-Dordogne. The war
starts. Dr\^ole de guerre. Going to work. Going to school. Fleeing
from village to village.  Playing cat and mouse. The second
landing. Return to Beaulieu. Return to Paris. Joining the
boyscouts. Learning languages. Israel becomes independent. Arrival in
Haifa. Kalay high school. Military training. The Hebrew Technion in
Haifa. Relativity. Asher Peres. Metallurgy. Return to France. Escape
from jail. Aviva.
\end{abstract}

\bigskip
\hfill I am the cat who walks by himself,\newline
\hspace*{\fill} and all places are alike to me.\newline
\hspace*{\fill} Rudyard Kipling\footnote{R. Kipling, {\it Just So
Stories\/}, MacMillan, London, 1902.}\bigskip

I am grateful to all those who contributed to this Festschrift which
celebrates my 70th birthday and therefore the beginning of my eighth
decade. In the Jewish religion, there is a prayer, ``she-hehhyanu'' to
thank the Lord for having kept us alive and let us reach this day. I
am an atheist and I have no Lord to thank, but I wish to thank many
other people who are no longer alive and who helped me reach this point.

\bigskip\centerline{\bf The city of lions}\medskip

First, I thank my parents, Salomon and Salomea Pressman, for leaving
Poland before World War~II and going to live temporarily in France,
so that we remained alive. Otherwise, I would not have been able to
celebrate my seventh birthday. My family originated in a city which
was called Lemberg when my parents were born in the Austrian empire,
Lw\'ow when I was born, Lviv (Ukraine) today. It also has a French
name (L\'eopol) and other names too. This means the city of lions
(in Russian, ``lev'' means lion). If it had a  Malay name, it would be
Singapore. It is the geographical center of Europe, half-way between
Rome and Moscow. You could live in that area in eight different
countries without ever moving from your home. The boundaries moved.

My father had always been a fighter. During World War~I, when he
was ten years old, as the city was encircled by Russian troops and
defended by the Austrians, there was a famine. The young boy went to
crawl between the two armies to unearth frozen potatoes and bring
them home to eat.  When he was 14, in 1919, he left home and went
to Palestine, then under British mandate, to build roads and drain
marshes. Three years later, he got the malaria and retuned home to
recuperate.  Then he had to serve in the Polish army and went to an
officer's course to reach the grade of ``aspirant.'' After that he went
to France to study electrical engineering, returned to Poland to marry,
and then again to  France where he had found work building power lines.

My mother came from a family of modern Jews. She spoke Polish
and German, not Yiddish like in my fathers' family which was more
traditional. Her father, Asher Schapira, was a journalist who was jailed
several times for expressing socialist opinions. He died when my mother
was 16. In the gymnasium where she learnt, one of her classmates was
Stanis\l aw (Stanley) Ulam. They were close friends and lost contact
only when World War~II started and Ulam was a professor of mathematics
in the USA. (Lw\'ow was a famous center for research in mathematics,
Stefan Banach taught there.) In 1933 my newlywed mother went to France
together with my father, and she followed him to remote places where
there was no electricity and he was building power lines.

\bigskip\centerline{\bf Beaulieu-sur-Dordogne}\medskip

And so it happens that I was born in a village, Beaulieu-sur-Dordogne,
in the Corr\`eze district, one of the poorest areas of France. But it
was more complicated than that. My mother had no previous experience
with childbirth and did not know what to do. The labor lasted three
days, and she was told that it was necessary to sacrifice the baby
or the mother. She answered: ``not him, not me.'' I am so grateful
to my mother for having been stubborn. After that, she could not
have other children. She told me that when I came out, I looked very
tired. Thereafter, whenever I complained to her that I was tired,
she answered ``you were born tired.''

My parents wanted to call me Asher after my deceased grandfather,
according to the Jewish tradition. However this was not a legal
first name in France. They called me Aristide, after a popular French
politician, Aristide Briand. That French name later saved
our lives, when we had to conceal our Jewish origin.

Soon after my birth, foreign workers and their families were
expelled from France because the economic depression had caused high
unemployment. My father returned to Poland to be unemployed there, and I
learnt to speak (and count, and write) in Polish. I was a happy Polish
child. In 1937, foreign workers were allowed to come again to France,
but without their families. My father found work in Paris and went to a
lawyer to declare me a French citizen (anyone born in France could
declare himself a Frenchman and the lawyer did that on my behalf). Then
the French consulate in Lw\'ow issued a French passport for me,
heavily stamped {\it non valable en Espagne\/} (not valid in Spain): at
that time, many young Frenchmen were volunteering to fight against
the Franco regime and the French government did not want me to join
that war. The passport was then also stamped with a Polish exit visa,
and with Czechoslovakian and German transit visas. The latter was
issued by {\it Deutsches Konsulat in Krakau\/}, with Gothic characters
and the usual Nazi symbols (swastika and eagle). An adult took me with
him in trains for two days, until we reached France on 19 October 1938
(no need of visa, since I had a French passport).

My father put me in a pension for children near Paris, and I quickly
forgot Polish and learnt French. My father came to see me every weekend
and asked the French authorities to let my mother join us (just taking
care of her French son was not considered as ``work''). It took many
months until my mother got a French visa, and finally she arrived six
weeks before the war broke out. I told the other children ``you see,
I also have a mother'' but I could not speak with her. I had forgotten
Polish, and she also had forgotten most of what she had known in French.

\bigskip\centerline{\bf The war starts}\medskip

We moved to my father's apartment in Paris, and then the war started. My
father, who had always been a fighter, volunteered to the Polish brigade
in the French army and told my mother that when we lose the war (there
was no doubt about the outcome) she should go to the small village
where I was born and people knew her, and if he could he would join us.

All adults in Paris were issued gas masks (children had no protection).
When I saw my mother trying her gas mask, I burst into tears. My mother
decided not to go to the shelter in case of air raid and to stay in
our apartment. We lived in a modern building and had six floors above us
to protect us.

In October 1939, my mother tried to send me to the neighborhood school,
but it refused to accept me because I was not yet 6 years old.  They had
to accept me in February 1940 so that I entered in the middle of the
year without any training in school discipline, and I was punished
several times for various mischiefs. The teacher once put me in the
garbage can.

\bigskip\centerline{\bf Dr\^ole de guerre}\medskip

The ``funny war'' ({\it la dr\^ole de guerre\/}) started in a rather
civilized way (at least in France). The French army was completely
demoralized. The Germans would send ahead a soldier speaking French, on
a motorcycle. When he met French troops he shouted: ``Halt! Lay down
your weapons! You are my prisoners.'' And they were prisoners. There
was no risk. In the worst case, the French officer would say: {\it
pardon\/}, it's an error, you are our prisoner. Not a single shot was
fired. (The war did not end so nicely, later I'll tell you about the
atrocities.) In June 1940, Paris was encircled by the German army and
declared ``open city'': it was not defended to avoid destruction
and the population was allowed to leave it in trains that crossed the
army lines. My mother and I took such a train and after three days of
zigzags arrived in Beaulieu, and rented there a room, the very room
where I was born.

Some time later, my father arrived in civilian clothes and with false
papers that an officer had obligingly given to him. Some Frenchmen
were really helpful people. We soon learnt that our radio had been
stolen from our apartment and was held by someone in a neighboring
village. My father went there and brought back the radio, a rare item
at that time, so that we could listen to the BBC and later learn of
the invasion of Russia and the war on other fronts.

\bigskip\centerline{\bf Going to work}\medskip

When I was 6 years old, I got my first job in the bakery of the
village. The task was to glue bread tickets on large sheets that
the baker would redeem for more flour. It was not a trivial task:
``if there is a sheet for tickets of 200 grams, you scatter here and
there a few tickets of 100 grams, as if these were errors.''  There
were also counterfeited tickets, not quite the same color. ``Don't
concentrate them in one area, but scatter them at random places, so
that they won't attract attention.'' When all the monthly sheets had
been filled, I was paid for my labour: a loaf of bread to bring home.
I am grateful to the baker who gave me my first practical education.

\bigskip\centerline{\bf Going to school}\medskip

In October 1940, I went to school in Beaulieu. My parents wanted me
to go to second grade. The teacher hesitated and gave me an exam:
I had to add and multiply numbers, and I recited the alphabet: ``a,
b\'e, c\'e, \ldots'' (not ``a, be, ce'' as the French say). I called
the letter j ``iott'' as in Polish. I passed the exam. Madame Salesse
was a clever teacher.

It was a small village, and a small school. Each teacher had to hold
two classes: first and second grade in one room, third and fouth
in another room, and so on. Their technique was the following: while
the teacher was taking care of one half of the children, the other half
had to do exercises and learn by themselves. In this way, I learnt to
learn. In that small school, there was a truly outstanding group of
teachers and I shall be grateful to them all my life. I realized how
great they were only after I returned to Paris.

These were the nicest years of my life. I was one of the children of the
village and felt perfectly at home. When I reached the third grade, there
was a problem: the teacher, Monsieur Lalite, was a prisoner of war and
a young woman had been hired to replace him. As she knew no one in the
village, she questionned the children. One of them was like me a Jewish
refugee. His name was Wolf. ``Where are you from?'' He was born in
Germany. ``Ah, the Germans are good soldiers!'' and then she dictated to
all the class: {\it Les Allemands sont de bons soldats\/}.

When she asked me the same question, I answered that I was born
here, in Beaulieu. That did not sound right. Typical names in the
Corr\`eze  district end in ``ac'' as in Jacques Chirac, the French
President. So she asked where were my parents from. I said from
Poland. ``The poor Poles, they have lost the war.'' (The poor French
had also lost the war.) Then she inquired about what the parents were
doing, and I answered that my father was cutting trees in the forest
(this was an understatement, he actually was the foreman of a group
of lumberjacks). That impressed her: ``You see that child, his father
has such a humble job and he is not ashamed of that. There are no silly
jobs, there are only silly people.'' And she dictated to the class the
French proverb: {\it Il n'est point de sot m\'etier, il n'est que de
sottes gens\/}. (Some time later she must have learnt that my father
had a degree in electrical engineering.)

Fortunately, the true teacher returned soon, under an exchange program
called {\it la rel\`eve\/}: young Frenchmen volunteered to work in
Germany in farms or factories and redeemed a prisoner of their
choice (and also released German workers to fight on other fronts).
After the war they were accused of collaborating with the enemy,
a thorny moral problem.

These were the happiest years of my life. I played with the other
children of the village, and like them went regularly to mass. I knew
that there were Jewish children who had taken refuge in the village
and I also played with them, but I did not know that I was one of
them. There was a piano in the house where we lived and my mother gave
piano lessons to little girls, but I had no patience for that. On my
8th birthday, I received as a gift my first book which was not a book
for the school: {\it Just So Stories\/}\footnote{{\it Histoires comme
\c{c}a\/}, in French.} by Rudyard Kipling.  It had a beautiful green
binding with golden letters, for sure it had been printed before the
war. There were wonderful stories in the book, in particular ``The
cat who walked by himself.'' I don't know what happened to the book,
whether I abandoned it in Beaulieu when we returned to Paris, or in
Paris when we left for Israel. I hope that other children enjoyed that
book. I bought a new copy, in English, when I was in London in 1965.

\bigskip{\centerline{\bf Fleeing from village to village}}\medskip

The good life ended in the spring of 1944, when someone discovered
that my father was Jewish and tried to blackmail him. My father, who
had always been a fighter, went to the woods and joined the partisans
(FFI) where he got the rank of lieutenant. My mother and I took refuge
in the house of a courageous French woman, Germaine Cheylac, who hid
us in her house for several weeks (so that we would not be arrested
in ours). I am most grateful to Miss Cheylac for being so helpful with
us. I still continued to go to school until my father told us to take
refuge further south, in a village called Nailloux from which one could
see the Pyrenees mountains on a clear day. People there knew my father
because he had also built power lines in that area before I was born. We
went to a family called Valette who owned a small hotel-restaurant and
already had some refugees in their hotel. There I also went to school,
but that village was smaller than Beaulieu and a single teacher had to
hold all the classes. We both quickly realized that I knew more than
him and he started learning things from me.

Then, German troops came to Nailloux. They did not come to fight, but to
rest from other fights. They took rooms in the village, flirted with the
French girls, and had for some weeks a good life. Many came to eat in
our restaurant. One of them once confided, when no other German could
hear him: ``Hitler kaputt.'' If an officer had known, it would have
been death penalty. Sometime later there was such a shortage of food
that people ate their cats, and sometimes rats too. There was a rich
supply of rats; some of them were so big that the cats would rather not
fight them. Monsieur Valette once took me for a trip through the
woods, to catch frogs that were then killed by putting them in boiling
water. Absolutely delicious, even if you are not hungry.  Another time,
we collected a large bag of snails, also quite tasty.  I decided to
learn German and bought a Berlitz book. The first lesson started {\it
Der Tee ist gut\/}, very useful for speaking with soldiers.

We listened to the BBC (this was forbidden of course) with
its encouraging slogans:\smallskip \\
\hspace*{2in}``The goal of fifty one nations,\\
\hspace*{2in}Capitulation without conditions.''\footnote{There was
also a recurring slogan that is hard to translate into English: {\it
Patience et courage, on les aura, les Boches!}}\smallskip

\noindent On June 6, there were great news: the landing in Normandy. At
once, everything changed. The partisans tried to prevent German
reinforcements to move North. Of course, they could not stop a regular
and much better equipped army, but they could harrass it and slow it
down. My father, who had always been a fighter, blew up a bridge under a
German military train, and was later decorated with {\it Croix de Guerre
avec Etoile de Bronze\/} (cross of war with a bronze star). The citation
read: while his group was under heavy fire and almost encircled,
he retreated only after having accomplished his mission, and brought
back all his men including two wounded ones, and all his equipment.

\bigskip{\centerline{\bf Playing cat and mouse}}\medskip

Partisans in the Nailloux region were less serious. One day they took
over the village and soon after them German troops entered and the
partisans fled away. The Germans were very angry. They burst into the
hotel and pointed their guns at us. Everyone raised hands. A soldier
disdainfully made a gesture to me to put down my hands. He told us:
there are plenty of dishes in the restaurant, the partisans have eaten
here (which was actually true, they must have been informed by our
neighbor, who was a notorious collaborator). Madame Valette calmly
answered that this was a restaurant and many people ate here. She
added that some time earlier there were even more dishes because German
soldiers had eaten here.  Meanwhile other soldiers searched the house
for weapons that would have been abandoned by the partisans.

Indeed, the partisans had left two guns in a back room. Fortunately,
one of the servants had seen the guns and thrown them in a ditch
outside the house.  That servant was considered as the ``idiot of the
village.'' (According to French tradition, in each village there is 
someone whose IQ is even lower than that of the other villagers, and
he is called the idiot of the village.) I don't remember the name of
that idiot\footnote{Maybe it was Jean-Fran\c{c}ois, as in the song of
Edith Piaf, {\it Les Trois Cloches.}} but I shall always be grateful
to him for his brilliant initiative that saved many lives, perhaps my
own. Maybe he was not really an idiot, he only pretented to be one.

Some days later, the partisans came again, arrested our neighbor the
collaborator and took him away. His daughter was hysterical and came
to our hotel to cry. Madame Valette brought her to a bed in a room
upstairs and people tried to comfort her. Naturally German soldiers
also came after the partisans had fled away. I was upstairs and looked
through the window, as any curious boy. I saw a soldier walking and
pointing his gun ahead of him. He also saw me and angrily shouted
at me, but I could not understand. Surely it was not {\it ``Der Tee
ist gut.''\/} Then the soldier briskly waved his hand, showing me to
get away from the window and inside the room: he did not want to hit
me if he had to open fire. I am grateful to this unknown German soldier
for having been concerned about my safety, even though I was his enemy.

\bigskip{\centerline{\bf The second landing}}\medskip

On 15 August 1944, Allied troops landed on the Mediterranean shore of
France and started moving north. On August 22, Paris was liberated, as
the French policemen turned their light arms against the Germans. The
German commander in Paris refused to obey the orders of Hitler which
had been to destroy the city. In the reception room of the hotel, we
had a large road map of France, and I had pinned on it small American
and British flags, to follow the advance of the armies as related
daily by the BBC.

The partisans came again and I pinned a small French flag on our
village Nailloux (I was later reprimanded for this act of patriotism).
Naturally, German troops soon arrived too and searched through the
hotel.  An elegantly dressed officer entered the reception room. He
saw the map and had a shock. Obviously, he did not listen to the BBC.
That was {\it verboten\/} to him. He looked at the map, and looked,
and looked, and looked, for a very long time.  And then he quickly
removed all my little flags, and stole the map. He was not angry at
us. We were not angry at him. Obviously, he needed a good map to find
his way out. We could always buy another road map in the local store,
and the little flags were ready.

\bigskip{\centerline{\bf Return to Beaulieu}}\medskip

Some time later, we could return to Beaulieu-sur-Dordogne. Nothing had
changed. Germans were never seen there. On the Dordogne river, there was
a long bridge, so important strategically that its pillars had places
ready for explosive charges, to blow up the bridge. The locations for
explosive charges were still clearly indicated. Miss Cheylac was so
happy to see us! I returned to school, in the class of Monsieur Faure.

However, further north, there had been the worst atrocities against
the French people during that war. In Oradour-sur-Glane, SS soldiers
from division ``Das Reich'' killed 642 villagers by herding them into
a church and burning the church. When Tulle, the capital of the
Corr\`eze district, was liberated, 99 men whom the SS had taken as
hostages were found hanged to lampposts and balconies.

In December, most of France was liberated and there were battles in the
Rhine valley, near the Siegfried fortification line that the Germans had
built many years earlier. The British soldiers had a song: ``We shall
hang our washing on the Siegfried line \ldots'' My father had enough
of fighting, and asked to be discharged from the FFI. We took a train
for Paris. It was a long trip, because many bridges had been blown up
and not yet repaired, and this was also the end of my good life.

\bigskip{\centerline{\bf Return to Paris}\medskip

In Paris, we returned to our apartment. By law, we had the right to
recover it, although we had not paid the rent during our absence. The
people who had used it (and paid the rent) had two weeks to find other
quarters. When we entered the apartment, we found that most of its
contents had been stolen, the furniture and rugs were damaged, but
still we were alive and at home.

I returned to the same neighborhood school that had accepted me for
a few months when I was six years old, but there was a difference. 
Now I was a country boy with a funny southern accent, but my level of
knowledge was far above that of the children in Paris. The headmaster
recommended to send me to a better school, and my parents enrolled me in
Lyc\'ee Voltaire, a long walk away (or three m\'etro stations).

There were other things I had to learn. Now I was Jewish, and need not
go to mass. On 8 May 1945, Germany capitulated (without conditions,
as promised) and the school gave us one day of vacation. Soon however,
there would be terrible news. No one remained alive in our family in
Poland (later we learnt that a child had been saved by a catholic
family which adopted him). My mother was so grieved that she lost
her mind and for two years could not function normally. My father
was stronger. He found work in a factory of electric appliances and
designed new ones so cleverly that they sold well. In his contract,
he was to receive a small percentage of the sales, and as this small
fraction became a lot of money, he was fired by his bosses who were
not content with their much larger fraction. Later he became a teacher
in an ORT vocational school.

Lyc\'ee Votaire was a large school and there were eight parallel
classes with slightly different programs, labelled A (Latin, Greek,
and one modern language), B (Latin and two modern languages), C (Latin
and one foreign language), and M (``modern'' program: no Latin, but
two foreign languages). The Lyc\'ee was in a lower middle class area
(near the P\`ere Lachaise cemetery) and still, only two classes were
``modern''.  Parents wanted their children to learn Latin, as
had been obligatory for everyone until 1940. Only then, following the
shock of German occupation, the French government took the courageous
decision of making Latin not mandatory.

The Lyc\'ee was not far from a Jewish area and about one quarter of
the pupils were Jewish and had lost her fathers in concentration camps.
The Jewish children laughed at me because I did not know Yiddish. Some
of the French children were antisemitic, a few of them virulently so.
One of them declared that all the Jews should go to Palestine (still
under British mandate). He had nothing against their religion, but he
loathed their behavior. Except that one Jewish child was perfectly OK.
You guess whom: a French educated child, who had recently learnt that he
was Jewish.

\bigskip\centerline{\bf Joining the boyscouts}\medskip

In the Lyc\'ee, another Jewish boy, Maurice Goldman, invited me to join
the boyscouts. There were in France four organizations of boyscouts:
Scouts de France (catholics), Eclaireurs Unionistes de France (EUF,
protestants), Eclaireurs Isra\"elites de France (EIF, Jewish), and
Eclaireurs de France (EDF, atheists). Of all the persons mentioned
above, Maurice is the only one with whom I still have contact. I met
him during a stay in the nuclear center of Saclay in 1956-57, and
again during a visit to Coll\`ege de France in 2001. His expertise
is NMR (that he calls RMN) and he is a member of the French Academy
of Sciences.

Another boyscout in our group is even more famous than him: Jacques
Benveniste, who was awarded {\it twice\/} the Ig-nobel Prize for
the discovery of ``water memory''\footnote{J. Benveniste et al.,
Nature {\bf333}, 816 (1988). See also editorial on page 787.} and
for its transmission to other water over the Internet. His father
was a physician and had a telephone, Tru~19-33, which we used for
transmitting urgent messages to others in our group.

In the EIF, I passed various exams to acquire badges and raise in the
ranks. I learnt the names of months and holidays in the Jewish calendar
(I still remember most of them), and I had my first festive Passover
dinner. We had a rabbi to teach us more of our religion. I often
antagonized him. When Rabbi Feuerwerker told us that it was forbidden
to mix milk and meat products in the same meal, I asked him why. He
said that it was written in the Torah, and I asked why we had to do
what was written in the Torah. The rabbi recommended to expel me from
the EIF, but my superiors refused. One evening he asked us to choose
a subject to teach us, and I proposed the life in Judea under Roman
occupation. He answered that nothing interesting happened at that
time. Obviously I should not have expected the learned rabbi to tell
us that a prominent social activist called Jesus was put to death by
the Roman occupation forces, in connivance with Jewish collaborators.
\clearpage
\bigskip\centerline{\bf Learning languages}\medskip

At that time I also started to learn Hebrew for my Bar-Mitsva ceremony
at age 13, where I had to publicly read a passage of the Torah (Exodus,
Chapt.~18). That passage included a conversation between Moses and
his father in law Jethro, the high priest of the Midianites, who
explained to Moses how to organize a hierarchy of public servants to
rule the people of Israel. I was already a rebel. I refused to chant
the text in the traditional fashion, and I read it in a natural way,
as a conversation. People in the synagogue were amazed and said that
this was the right way of reading the Torah.

I also learnt two foreign languages in Lyc\'ee Voltaire. Nearly
everyone took English as the first foreign language. For the
second one, the choice was between German and Spanish. My father
forbade me to learn German. He also forbade me to ever have any
contacts with German people (though he carefully added: not with
the present generation).  I still remember some Spanish, but by now
most of it is {\it olvidado\/}.  Starting to learn two new languages
simultaneously may cause confusion. Once I said ``al'' instead of
``sobre'' (over). Most pupils took German as their other language,
including the Jewish ones who had lost their fathers. It was easier
for them, because of the resemblance with Yiddish.

\bigskip\centerline{\bf Israel becomes independent}\medskip

On 15 May 1948 Israel was declared an independent state and there
was a great joy in the Jewish community. We applied to the Israeli
Consulate for an immigration visa. At that time there was not yet in
Israel a ``law of return'' which guarantees the right of immigration
to every Jew.  The Consulate refused: ``You have an apartment in
Paris. There are more than a million Jews in the camps, they have to
come first.'' We came more than a year later.

Just before we left Paris, there was Yom Kippur (the holiest
day in the Hebrew calendar) and I went to attend services in the
neighborhood synagogue. There was a mixed congregation, with people
from Eastern Europe who spoke Yiddish, and people from North Africa
who did not. The rabbi started a sermon to explain that now there was
a State of Israel that we had to support. As he spoke in French, the
East Europeans shouted at him ``Reth Yiddish!'' He answered that not
everyone understood that language, and several people confirmed that
this was indeed true. The rabbi explained that he spoke the language of
the country where we lived, so that everyone should understand. This
did not help. The East Europeans continued to shout at him to prevent
him from speaking. Then the rabbi got very angry, and shouted back
{\it ``Sales juifs!''\/} (filthy Jews). That was enough for me. I got
up and left the synagogue, and never since then did I set foot in a
synagogue again.

To leave for Israel, I needed a new French passport. I was then 15
years old, and from my picture in that passport it appears that I
was a handsome young man. Indeed, all the girls were running after
me. But I was extremely shy, and I was running away from them.

\bigskip\centerline{\bf Arrival in Haifa}\medskip

We took a special train ({\it le train juif\/}, as the railroad
employees called it) and arrived in Marseille. For the first time
in my life, I saw the sea. We took the ship Negba. The sea was calm
and after a few days we reached the Holy Land. We were filled with
emotion: at last we are at home! We stayed overnight in the ship,
and went ashore the following morning. I shall never forget the welcome.

First, I was doused with a spray of DDT on my head (at that time I
had plenty of hair). Then we were bussed to an abandoned British army
camp. In each barrack, there were 60 beds. Our hosts didn't even
charge us the rent. I had never expected such an hospitality.

My father was not unemployed a single day. He immediately found work.
You guess what: building power lines. So he could rent a room near
Tel-Aviv and soon I could go to school.

\bigskip\centerline{\bf Kalay high school}\medskip

I enrolled in the Kalay high school in Givatayim, which was run by the
workers union ({\it histadruth\/}). At that time, schools in Israel
were run by political organizations, as they had been during the
British mandate before independence. That school had orders to accept
new immigrants, even if they didn't know Hebrew, and to give them a 10\%
discount on tuition fees.

Tuition fees? That was a new notion for me. In the French Republic
where I was born, which is the land of freedom, education is free:
tuition free and religion free (that is, anti\-religious). Now I
had to learn many books of the Bible --- only the Old Testament of
course. Since this was a socialist school, the Bible was not taught as
a religious book, but as an expression of the struggle of classes. The
Bible teacher also taught us the precepts of Karl Marx.

The pupils in that school were not supposed to go on to university
studies but to start new kibbutzim (collective farms). In addition to
the mandatory studies in Bible, Hebrew, Maths and English,
there were also courses in history, biology and Arabic. I had other
plans. I asked to be exempted from Arabic, because I knew French
that the school did not teach. The management agreed, and when they
gave me my final certificate two years later, I had a mark 10/10
in French. The ministry of education refused to accept it because the
school was not accredited for teaching French. The school then gave
me a passing grade 6/10 in Arabic.

I also did not want to learn history or biology, but physics and
chemistry that the school did not teach at all. I went to the ministry
of education and proposed to pass my matriculation exams as an external
student, unrelated to any school. However, that was permitted only to
people who did not attend any school. They found a solution: I would be
examined with the pupils of a neighboring school in Ramat Gan (which
belonged to a right wing party). That school kindly allowed me to
practice in their chemistry lab. I had to learn the rest from books. For
Marxism, there was no solution and I had to take the exam in my school.

I passed the written exams in physics and chemistry in the Ramat Gan
school. My grade in physics was 8/10, because my book used cgs units,
and the exam was in MKS.

\bigskip\centerline{\bf Military training}\medskip

I matriculated and got a deferment from military service, but I
still had to do two months of basic army training during the summer.
I was not a very good soldier. I broke my glasses, and the military
optometrist who prescribed new ones said that my vision was so acute
that I would find the most beautiful girl in the world to marry. (How
did he know so long in advance?)

Together with me in the army there were two French immigrants who had
served for one year in the French army in Indochina and were already
very well trained. They had to do again their military service in
Israel. (Only later, in 1961, France and Israel made an agreement
to prevent duplicating military services, and I went to the French
Consulate in Haifa to regularize my situation.)

\bigskip\centerline{{\bf The Hebrew Technion in Haifa}\footnote{Today
called Technion --- Israel Institute of Technology.}}\medskip

I enrolled in Technion in the department of mechanical engineering.
I really wanted to learn physics, but my father had warned me that I
would never find a job. (I am in good company: Eugene Wigner learnt
chemical physics because his father had told him that he would not
find a job as a physicist.)

Some weeks before I started my studies, as I had nothing to do except
giving private lessons, I learnt from a mathematics book of my father
and solved all 2500 exercises. After that, I used not to come to
lectures, but to learn by myself, at my own pace. I would still come
to one lecture by each teacher, to see who he was. When I came to a
lecture by Professor Zakon, the teacher of mathematics, I caught an
error in a proof. He was terrified, and promised to give me the grade
100 provided that I never come to his lectures.

In this way, I acquired a good reputation, got top grades sometimes
without exam, and had plenty of time left to learn physics. Only
when I reached the fourth year I had difficulties with the one of the
teachers who lived in Tel-Aviv and did not know that I had a
good reputation.

\bigskip\centerline{\bf Relativity}\medskip

I learnt the elements of relativity theory from Einstein's booklet
{\it The Meaning of Relativity\/}, and stumbled on a difficulty:
what were the Lorentz transformation laws of electric and magnetic
susceptibilies. I asked Prof.\ Nathan Robinson, the head of the physics
laboratory for freshman, whom I knew. He had studied many years earlier
in the University of Berlin, and in his student notebook he had the
signatures of his teachers, including Einstein and Planck. Meanwhile,
however, he had specialized in rainfall measurements and had no idea of
what I asked. He sent me to ask a new professor who had just arrived
at Technion: Nathan Rosen who had been a close collaborator to Einstein.

Rosen also did not know the answer to my question (later I found
it in Landau and Lifshitz, {\it Electro\-dynamics of Continuous
Media\/}). Rosen lent me his personal copy of Tolman's book {\it
Relativity, Thermodynamics, and Cosmology\/}, from which I could learn
more. Then sometime during my second year of mechanical engineering,
as I was supposed to learn about bolts and nuts, I found that the
Maxwell equations in curved space had a property that the resulting
wave equation was like that for particles of finite mass (the Compton
wavelength was equal to the radius of curvature of spacetime).

I wrote a short note and sent it to Prince Louis de
Broglie,\footnote{Nobel Prize in Physics, 1929.} who was the perpetual
secretary of the French Acad\'emie des Sciences. De Broglie kindly
anwered that Madame Tonnelat had found a similar result and I should
quote her work. I did that and he accepted my note, which was published
in {\it Comptes Rendus\/} {\bf 239}, 1023 (1954). A copy of his
handwritten letter to me appears here.

I am most grateful
to Prince Louis de Broglie. He was not only a prince by birth. He also
was a noble person who understood that beginners should be encouraged
in their first steps, even if their work had little value.

\bigskip\centerline{\bf Asher Peres}\medskip

When I came to Israel, I informally restored my true first name Asher.
However my formal documents (in the army, and my identity card) still
had my French first name Aristide. Once my army commander called me
``Artistide.'' Also I had difficulties to cash checks bearing the name
Asher. I decided to legalize the situation, and also to change my family
name Pressman to a Hebrew name, as was then customary in Israel. Most
people tried to keep in the Hebrew name someting reminding the old
one. There were plenty of jokes. Someone would change his name from
Bernstein to Ben-Satan (son of Satan).

The government agency who took care of name changes also gave advice on
this matter. They recommended to shorten Pressman into Peres (a mountain
eagle, {\it gypaetus barbatus\/}), or to keep all the consonants and
become Afarsamon\footnote{In Hebrew, the same character is used for f
and p (and ph). Most vowels are not written.} (a fruit, {\it diospiros
virsinata\/}).  I chose Peres, and my parents reluctantly followed suit.
That change of name later helped to save me from jail when I returned
to France for one year of studies in 1956.

With hindsight, I should have chosen Afarsamon, so as to be the first
author of my future publications. For example the famous teleportation
paper (1993) would have been Afarsamon et al., instead of Bennett
et al..

\bigskip\centerline{\bf Metallurgy}\medskip

In my third year I had a course in metallurgy. As usual, I went to have
a look at the teacher, Professor Taub, and I read a book: Cottrell,
{\it The Physics of Metals\/}. When the exam came, I answered with
notions well beyond what the professor had taught. He called me and
proposed that I become his teaching assistant for the following year,
or else do some research. You guess what I chose. The project I had
in mind was to measure the cold work energy in copper: when a metal
is subjected to plastic deformation, most of the work done is released
into heat, but a small part remains in the metal as internal stresses.
The difference can be measured by calorimetric methods. I proposed a
direct measurement by making an electric cell whose electrodes would
be copper under stress and unstrained copper. I estimated the potential
difference to several millivolts.

This was a reckless idea. I did not know that electrochemistry
is akin to black magic, with unexplained surface phenomena called
``polarization'' and other parasitic effects much larger that what I
intended to measure. I saw the potential difference change erratically
with time. To stabilize it, I introduced a third electrode (a grid,
as in a triode) and let an alternating current pass which was not
recorded by the DC voltmeter. When I finally obtained reproducible
results, whatever they meant, Professor Taub let me publish
them.\footnote{A. Peres, Bull. Research Council Israel {\bf 6C}, 9
(1957).} This was my first and last experimental paper. Thereafter I
worked only on theory, which is much safer.

\bigskip\centerline{\bf Return to France}\medskip

At that time, the French and Israeli governments agreed that the French
would help build a powerful nuclear reactor in southern Israel (its
purpose was not disclosed, but it was obvious). Israel needed to train
a large number of experts for its maintenance, and Technion was asked
to send a metallurgy expert for training in France. Professor Taub
recommended to send me although I still was an undergraduate, because
I knew French. This assignment was considered as part of the military
service that I had to start upon graduation.

I obtained my first Israeli passport (a nontrivial task at that time,
because the government did not like Israelis to travel) and I went to
the French Consulate in Haifa to get a visa to study for one year. The
Consulate gave me a visa for three months only, specifying that it
could not be extended under any circumstances. I arrived in Paris in
October 1956 and asked the scientific attach\'e in the Israel Embassy,
Mr.~Shalevet Freier\footnote{Freir later became the head of the Israel
Atomic Energy Commission.} to help me. He put me in contact with a
special police agent called Monsieur Cirinelli, who accompanied me
three times to Pr\'efecture de Police to extend my visa.

I found a room with an old lady, Madame Grimon, who had an apartment
in Ile Saint Louis, the smaller of the two islands that had been the
ancient Paris (the larger island has the Notre-Dame cathedral). She
rented two rooms in her apartment. Her other tenant was Monsieur
Goubault who had been a pilot in the French Air Force during World
War~I and meanwhile had become an expert on wines. Ile Saint Louis
is charming. Once I asked in the street where could I buy a small
lock and was answered ``you won't find that in {\it the\/} island,
you must go to Paris.''

I started my studies in Centre d'Etudes Nucl\'eaires de Saclay. My old
friend Maurice Goldman was also there, working on NMR with Anatole
Abragam.  First, I had a course on nuclear reactor structure which
gave me a formal degree in ``atomic engineering,'' and then I went for
a stage in the metallurgy department. A security guard there saw in
my foreign identity card that I was born in Beaulieu.  He was born in
Uzerche, also in the Corr\`eze district. For him I was {\it un pays\/}
(a fellow countryman) and he liked me.

\bigskip\centerline{\bf Escape from jail}\medskip

My visits to Pr\'efecture de Police with Monsieur Cirinelli arose
suspicion. As my Israeli passport showed, I was born in France. If I
also had French citizenship, I should have done my military service
there, or else I had been condemned {\it in absentia\/} to one year in
prison, and after it still had to do the regular military service in
France. Fortunately, the French police is highly compartmented and the
branch that keeps a watch on foreigners is not the one that looks for
deserters, because these are two disjoint sets. However, I belonged to
an even smaller subset: foreigners born in a remote village in rural
France. One day, I received a letter requesting to bring my birth
certificate to the nearby police station. That was no problem. I
asked my former teacher Madame Salesse in Beaulieu to send me a birth
certificate specifying that my parents had Polish citizenship, and
all was quiet for some time.

However in September I got another letter, ominously carrying both my
old and new names, asking me to report again to the police station. The
nooze was tightening. I reported instead to the El-Al office, and asked
to take the first plane for Israel. This was just before the Hebrew New
Year and the other high holidays. All seats were booked for several
weeks, but I knew personally the El-Al clerk, Suzanne Puderbeutel. We
had played together when we were small children and our parents lived
in the same building. Recently we had gone once to the movies. I am
grateful to Suzanne for giving me a seat in the first plane, so that
I could fly to safety.

After Suzanne, there is no one else I have to thank for ``having kept
me alive and let me reach this day.'' At last I was safe.

\bigskip\centerline{\bf Aviva}\medskip

I rented an apartment in Haifa (one room without kitchen) and I started
to teach nuclear engineering in Technion. This was considered as part
of my military service in Israel. I also started graduate work toward
a PhD degree in physics under the guidance of Professor Rosen.

On 5 January 1958, I took a train from Tel-Aviv to Haifa. Two pretty
young women wearing army uniforms took seats opposite to me and started
speaking French. One of them was indeed {\it very\/} pretty. I had
always been terribly shy with women, but they could not know that I
was shy. I decided to be courageous and I bravely entered into their
conversation, in French.

Two hours later, when the train arrived, Aviva and I exchanged
addresses. We already wanted to marry, but we did not reveal
that to each other until some weeks later. We actually married in
August.\medskip

The rest of my story is in my formal CV.
\end{document}